\documentclass[12pt,preprint,psfig]{aastex}

\newcommand{\eg}{{\it e.g.}\,}
\newcommand{\ie}{{\it i.e.}\,}
\newcommand{\ea}{et al.\,}
\newcommand{\be}{\begin{equation}}
\newcommand{\ee}{\end{equation}}

\def\t{\theta}
\def\cc{{\rm cm}^{-3}}
\def\g{\gamma}

\def\sy{{\rm synch}}
\def\scat{{\rm scat}}
\def\esc{{\rm esc}}

\def\tr{{\rm cross}}
\def\ic{{\rm IC}}
\def\br{{\rm brem}}

\def\crit{{\rm cr}}

\def\min{{\rm min}}
\def\max{{\rm max}}
\def\tot{{\rm total}}

\def\mug{\mu{\rm G}}

\shorttitle{RXTE Observation of 1E0657-56}
\shortauthors{Petrosian, Madejski \& Luli}

\begin{document}


\title{Analysis and Interpretation of Hard X-ray Emission from
the Bullet Cluster (1E0657-56), the Most Distant Cluster of Galaxies Observed
by {\it RXTE}}


\author{Vah\'{e} Petrosian\altaffilmark{1,2,4}, Greg Madejski\altaffilmark{2,3} and Kevin Luli\altaffilmark{1}}
%\affil{Kavli Institute for Particle Astrophysics and Cosmology and Department of Physics, %Stanford University, Stanford, CA 94305}


\altaffiltext{1}{Department of Physics, Stanford University, Stanford, CA, 94305 email: vahe@astronomy.stanford.edu}
\altaffiltext{2}{Kavli Institute for Particle Astrophysics and Cosmology, Stanford University, Stanford, CA 94305}
\altaffiltext{3}{Stanford Linear Accelerator Center, Menlo Park, CA 94025}
\altaffiltext{4}{Department of Applied Physics, Stanford University, Stanford, CA, 94305}


\begin{abstract}
Evidence for non-thermal activity in clusters of galaxies is well established
from radio observations of synchrotron emission by relativistic electrons.
New windows in the Extreme Ultraviolet and Hard X-ray ranges have provided 
for more powerful tools for the investigation of this phenomenon.
Detection of hard X-rays in the 20 to
100 keV range have been reported from several clusters of galaxies, notably
from Coma and others. Based on these earlier observations 
we identified the relatively high redshift cluster 1E0657-56 
(also known as RX J0658-5557) 
as a good candidate for hard
X-ray observations. This cluster, also known as the bullet cluster, has many
other interesting and unusual features, most notably that it is undergoing 
a merger, clearly visible in the X-ray images. Here we present results from a
successful \textit{RXTE} observations of this cluster.  
We summarize past observations
and their theoretical interpretation which guided us in the selection process.
We describe the new observations and 
present the constraints we can set on the flux
and spectrum of the hard X-rays. Finally we discuss the constraints
one can set on the characteristics of
accelerated electrons which produce the hard X-rays and the radio radiation.
\end{abstract}


%% subject header.  Objects should be in the appropriate "individual"
%% headers (e.g. quasars: individual, stars: individual, etc.) with the
%% additional provision that the total number of headers, including each
%% individual object, not exceed six.  The \objectname{} macro, and its
%% alias \object{}, is used to mark each object.  The macro takes the object
%% name as its primary argument.  This name will appear in the paper
%% and serve as the link's anchor in the electronic edition if the name
%% is recognized by the data centers.  The macro also takes an optional
%% argument in parentheses in cases where the data center identification
%% differs from what is to be printed in the paper.


%\include{introd}
\section{INTRODUCTION}
\label{intro}

The intra-cluster media (ICM) of several clusters of galaxies, in addition to
the well studied thermal bremsstrahlung (TB) emission 
dominating in the $\sim 2 - 10$ keV soft
X-ray (SXR) region, show growing evidence for non-thermal activity. 
This activity was
first observed in a form of {\it diffuse radio} 
radiation (classified either as relic 
or halo sources). The first cluster with diffuse emission detected 
in the radio band 
was Coma, and recent systematic searches have identified more than 40 clusters 
with halo or relic sources.  In the case of
Coma, the radio spectrum may be represented by a broken power law (Rephaeli
1979), or a power law with a rapid steepening (Thierbach \ea 2003) or 
with an exponential cutoff 
(Schlickeiser \ea 1987).  There is little doubt that this radiation 
is due to synchrotron 
emission by a population of relativistic electrons 
with similar spectra, however,
from radio observations alone one cannot determine the energy of the
electrons or the strength of the magnetic field.  Additional observations or
assumptions are required.  Minimum total (particles plus field) energy or 
equipartition assumptions imply magnetic field strength 
of $B\sim \mu{\rm G}$, in rough agreement
with the Faraday rotation measurements (\eg Kim \ea 1990), and a population of
relativistic electrons with Lorentz factor $\g \sim 10^4$.
In the  papers cited above, it was also realized 
that because the  energy density of
the Cosmic Microwave Background (CMB) radiation (temperature $T_0$)
$u_{\rm CMB}= 4\times 10^{-13}(T_0/2.8 \, {\rm K})^{4}\,{\rm erg} \,\cc$ 
is larger than the  magnetic energy density 
$u_{\rm B}= 3\times 10^{-14}(B/\mu{\rm G})^2 \, {\rm erg} \,\cc$, 
most of the energy of the relativistic electrons will be radiated via inverse
Compton (IC) scattering of the CMB photons, producing 
a broad photon spectrum (similar to that observed in the radio band) around 
50 keV (for $\g \sim 10^4$). 
Thus, one expects a higher flux of non-thermal 
X-ray radiation than radio radiation.
Detection of HXRs would then break the degeneracy 
and allow determination of the
magnetic field and the energy of the radiating electrons. Moreover, since the 
redshift dependence of the CMB photons is known, then in principle, 
the cosmological evolution of these
quantities can also be investigated.

While the detectability of such radiation would be 
easiest in the soft X-ray range, 
where very sensitive imaging instruments are available, 
in reality, in this range, it is masked by the 
prominent thermal bremsstrahlung emission, a general 
characteristic of clusters.  
With this, the most promising is either the hard X-ray 
(HXR) band, beyond the energies where 
TB flux dominates, or alternatively, in the very 
soft X-ray or extreme 
ultraviolet regime.  
Recently  HXR emission (in the 20 to 80 keV range) 
at significant levels above that expected 
from the thermal gas was detected by instruments on board 
{\it Beppo}SAX and \textit{ RXTE } 
satellites from Coma (Rephaeli \ea 1999, 
Fusco-Femiano \ea 1999, Rephaeli \& Gruber 2002, Fusco-Femiano \ea 2004%
\footnote{The results of this paper have been challenged by an 
analysis performed with different software by  Rossetti \& Molendi (2004).}), 
Abell 2319 (Gruber \& Rephaeli 2002), 
Abell 2256 (Fusco-Femiano \ea 2000, Rephaeli \& Gruber 2003, and 
Fusco-Femiano, Landi, \& Orlandini 2005), 
and  a marginal ($\sim3\sigma$) detection from Abell 754 
and an upper limit on Abell 119 (Fusco-Femiano \ea 2003).  
We also note that a possible recent detection of
non-thermal X-rays, albeit at lower energies, has been reported from a
\underline{poor} cluster IC 1262 by Hudson et al. (2003).  All 
those clusters are in the redshift range
$0.023<z<0.056$.  
Notable recent exception at a higher redshift 
is Abell 2163 (Rephaeli, Gruber, \& Arieli 2006)
where the reported nonthermal flux is consistent with the 
upper limit set by {\it Beppo}SAX (Feretti 
\ea 2001).
It should also be noted that excess radiation was
detected in the 0.1 to 0.4 keV band by {\it Rosat, Beppo}SAX and 
{\it XMM-Newton} and in the EUV region (0.07 to 0.14 keV) similar 
excess radiation was detected by the {\it Extreme
Ultraviolet Explorer} from Coma (Lieu et al.  1996) and some other clusters.  
A cooler ($kT\sim 2$ keV) component and IC scattering of
CMB photons by lower energy ($\g \sim 10^3$)
electrons are two possible ways of producing this excess radiation.  
However, some of the observations and the emission process 
are still controversial (see Bowyer 2003).

Here we present results from $\sim 309$ ks \textit{RXTE } observations of
cluster 1E0657-56 (also known as RX J0658-5557) with a considerably
higher redshift of $z=0.296$, which manifests many other interesting features
(Markevitch 2005).  All those hard X-ray observations - especially with the 
addition of 1E0657-56 - encompass a wide range of temperature, 
redshift and luminosity, indicating that HXR emission may be a common property 
of all clusters with significant diffuse radio emission.  
In the next section we give brief descriptions of the emission processes and
the considerations which led to the selection of 1E0657-56 as a target for
HXR observations by \textit{RXTE }. In \S 3 we describe the observations 
and the results from our spectral fits. Finally in \S 4 we discuss the 
significance of our results and their implication for the radiation 
and acceleration mechanisms.  

\section{EMISSION PROCESSES AND TARGET SELECTION}
\label{select}

As stated above electrons of similar energies can be responsible for both the
IC$-$HXR
and synchrotron$-$radio emission and the ratio of these 
fluxes depends primarily
on the ratio of the  photon (CMB in our case) to magnetic 
field energy densities.
For a population of relativistic electrons with the spectrum
$N(\g)=K\g^{-p}=N_\tot (p-1)\g^{-p}\g_\min^{p-1}; {\rm for} \, \g>\g_\min$, the 
spectrum of the emitted luminosities of both radiation components is given by

\be\label{nufnu}
\nu L_i(\nu)=cr_0^2N_\tot \g_\min^{p-1} u_i A_i(p)(\nu/\nu_{\crit,i})^{2-\Gamma},
\ee
where $r_0$ is the classical electron radius, $c$ is the speed 
of light and $\Gamma=(p+1)/2$ is the photon number spectral index% 
\footnote{These expressions are valid for $p>3$ or $\Gamma>2$. For smaller 
indices an upper energy limit $\g_\max$ also must be specified and the 
above expressions must be modified by other factors which are 
omitted here for the sake of simplification.}.

For synchrotron
$\nu_{\crit, \sy} = 3\g_\min^2\nu_{B_\perp}/2$, 
with $\nu_{B_\perp}=eB_\perp/(2\pi mc)$
and $u_\sy=B^2/(8\pi)$,
and for IC, $u_i$ is the soft photon energy density and 
$\nu_{\crit,\ic} = \g_\min^2\langle h\nu \rangle$. For black body photons
$u_\ic = (8\pi^5/15)(kT)^4/(hc)^3$ and $\langle h\nu \rangle=2.8 \, kT$. 
$A_i$ are some simple functions of the electron index $p$ and are of the
order of unity
(see \eg Rybicki \& Lightman 1979). Because we know the temperature of the CMB
photons,
from the observed ratio of fluxes we can determine the strength of the
magnetic field.
For Coma, this requires the volume averaged magnetic field to be 
${\bar B}\sim 0.1 \, \mug$, 
while equipartition gives ${\bar B}\sim 0.4 \, \mug$
while Faraday rotation measurements give the average line-of-sight field of 
${\bar B}_{\rm l}\sim 3 \, \mug$ 
(Giovannini et al.  1993, Kim \ea 1990, Clarke \ea  2001;  2003).
(In general the Faraday rotation measurements of most clusters
give $B> \mug$; see \eg Govoni et al. 2003.)  
However, there are several factors which may
resolve this discrepancy. Firstly, the last value assumes a chaotic magnetic 
field with scale of few kpc which is not a directly measured quantity 
(see \eg Carilli \& Taylor 2002).  Secondly, the accuracy of these results 
have been questioned by Rudnick \& Blundell (2003) and defended by 
Govoni \& Feretti (2004) and others. Thirdly, as pointed by 
Brunetti \ea (2001), a strong gradient in 
the magnetic field can reconcile the difference between the volume and 
line-of-sight averaged measurements.  
Finally, as pointed out by Petrosian 2001 
(P01 for short), this discrepancy can be alleviated by
a more realistic electron spectral distribution (\eg the spectrum
with exponential cutoff suggested by Schlickeiser \ea 1987) and/or
a non-isotropic pitch angle
distribution.   In addition, for a population
of clusters  observational selection
effects come into play and may favor Faraday rotation detection
in high $B$ clusters which will have
a weaker IC flux relative to synchrotron.

A second possibility is that the HXR radiation is produced via
bremsstrahlung  by a population of electrons with
energies comparable and larger than the HXR photons. 
If a thermal distribution of electrons is the source of this radiation, such 
electrons must have a much higher temperature
than the gas responsible for the SXR emission. 
For production of HXR flux up to 50 keV
this requires a gas with $kT>30$ keV  and 
(for Coma) with an emission measure about 10\% of
that of the SXR producing plasma. Heating and 
maintaining of the plasma to such high
temperatures in view of rapid equilibration expected 
by classical Spitzer conduction is problematical.
It has also been suggested by various authors
(see, \eg En\ss lin, Lieu, \& Biermann 1999; Blasi 2000)
that the HXR radiation is due to non-thermal bremsstrahlung (NTB) by a
power law distribution of
electrons in the 10 to 100 keV range. However, it was shown in P01 that the
NTB process faces a serious difficulty, which is  hard to circumvent,
because
compared to Coulomb losses the bremsstrahlung yield is very small; $Y_{\br}
\sim
3\times10^{-6}(E/25 \, {\rm keV})^{3/2}$ (see Petrosian 1973).  
Thus, for continuing 
production of a HXR
luminosity of $4\times 10^{43}$ erg s$^{-1}$ (observed for Coma), 
a power of $L_{\rm
HXR}/Y_{\br} \sim 10^{49}$ erg s$^{-1}$ must be continuously fed into 
the ICM, increasing its
temperature to $T\sim 10^8$ K after $3\times 10^7$ yr, or to $10^{10}$ K in a
Hubble time%
\footnote{These estimates are based on energy losses of electrons in a cold
plasma  which is an excellent approximation for electron energies
$E\gg kT$. As $E$ nears $kT$ the rate of loss of energy decreases and the
bremsstrahlung yield increases. For $E/kT>4$ this increase will be
at most about
a factor of 2.}. Therefore, the NTB emission phase must be very short
lived. A possible way to circumvent the rapid cooling of the hotter plasma by
conduction or rapid energy loss of the non-thermal particles is to 
physically separate
these from the cooler ICM gas. Exactly how this can be done
is difficult to determine but strong
magnetic fields or turbulence may be able to produce such a situation.

In what follows we shall assume the IC process as the 
working hypothesis for the origin of the HXR flux.
However, in view of the above mentioned difficulties more observations are 
acutely needed to determine the HXR emission process.  
Such observations are challenging:  thermal emission from clusters of galaxies 
can be well characterized as thermal bremsstrahlung with 
emission lines due to atomic 
transitions, with a characteristic temperature of 
a $\sim$ a few to $\sim 15$ keV.  
This means that any reliable detection of the non-thermal 
component relies on an instrument sensitive 
above at least 10 keV.  Sensitive measurements require 
instruments employing focusing optics, 
but no such instruments are currently operational.  We must thus rely on the 
collimated detectors on-board of {\it Beppo}SAX, \textit{RXTE}, 
or \textit{Suzaku } HXD, but for such instruments, 
good limiting sensitivity requires long observations.  
Therefore, the target selection - based on the good prediction of the 
HXR flux - requires careful
considerations of all aspects of the phenomenon.  

The most important {\it selection criterion}
is the presence of diffuse radio emission.
About 40 clusters are known to have such emission, which can be classified as
halo or relic (see Giovannini, Tordi, \& Feretti 1999, Giovannini \&
Feretti 2000, Feretti \ea  2000).  The fraction of clusters with
diffuse radio emission rises with the SXR luminosity (Giovannini \ea
 1999) which is in turn correlated with temperature, via $L_{\rm SXR}\propto
T^2$ (see e.g.  Allen \& Fabian 1998), indicating a correlation of the
non-thermal activity with temperature.  In addition, turbulence and shocks,
present in merging clusters, are the most likely agents of acceleration
and could be the cause of the higher temperatures, as well as the nonthermal 
activity. 
Consequently, high temperatures
and presence of substructure also must influence the selection  (see
Buote 2001 and Schuecker et al.  2001). Note that higher temperatures
make the detection of the HXR flux more
difficult. However, at higher redshifts
this effect is offset by the spectral redshift $z\equiv Z-1$. 
Based on the criteria given above, 
a list of some of the most promising clusters are given in Table 1.

\vspace{0.5in}

\begin{center}
\center{{\large {\bf Table 1}}}
\vspace{-0.16in} \center{\bf OBSERVED AND ESTIMATED PROPERTIES OF CLUSTERS}

\begin{tabular} {|l|ccccc|cc|}
\hline \hline 
Cluster & $z$ & $kT^a$ & $F_{1.4 {\rm GHz}}^b$ & $\t^{c,b}$ & $F_{\rm SXR}$ 
& $B^d$ & $F_{\rm HXR}^e$ 
\\
 & & keV & mJy  & arcmin & $F_0^f$ & $\mug$ & $F_0^f$ 
\\ 
\hline 
Coma   & 0.023 & 7.9 & 52 & 30 & 33 & 0.40 & 1.4(1.6) 
\\ 
A 2256 & 0.058 & 7.5 &400 & 12 & 5.1 & 1.1 & 1.8(1.0) 
\\
1E0657-56 & 0.296 & 15.6 & 78 & 5 & 3.9 & 1.2 & 0.52(0.5) 
\\
A 2219       & 0.226 &12.4 & 81 & 8 & 2.4 & 0.86 & 1.0                       
\\ 
MACSJ0717    & 0.550 & 13 & 220 & 3 & 3.5 & 2.6 & 0.76
\\
A 2163       & 0.208 & 13.8 & 55 & 6 & 3.3 & 0.97 & 0.51 
\\ 
A 2744       & 0.308 & 11.0 & 38 & 5 & 0.76 & 1.0 & 0.41
\\ 
A 1914       & 0.171 & 10.7 & 50 & 4 & 1.8 & 1.3  & 0.22
%\\
%A 520 & 0.203 & 8.3 & 38 & 1.3 & 4 & 0.9 & 0.30 &6
%\\ 
%A 665 & 0.182 & 9.0 & 31 & 1.1 & 5 & 1.2 & 0.0.2 &3 
%\\ 
%A 2255 & 0.081 & 7.3 & 45 & 1.3 & 5 & 1.8 & 0.25 &4 
\\ 
\hline \hline
\end{tabular}
%\vspace{0.15in}
\end{center}
%\hspace{0.6in}
\noindent $^a$ From Allen \& Fabian (1998), except 1E0657-56 data from Liang \ea (2000)\\
\noindent $^b$  From Giovannini \ea (1999, 2000), 
except 1E0657-56 data from Liang \ea (2000)\\
%\vspace{-0.15in}
\noindent
%\hspace{0.6in}
\noindent $^c$ Approximate largest angular extent.  \\
\noindent $^d$ Estimates based on equipartition. \\
%\vspace{-0.15in}
\noindent $^e$ Estimates assuming $\zeta\g_\min=10^6$, with observed values in parentheses for Coma 
from Rephaeli \ea (1999; 2002), and Fusco-Femiano \ea (1999; 2004) 
and for Abell 2256 by Fusco-Femiano \ea (2000;  2005) and Rephaeli \& Gruber (2003).\\
%\vspace{-0.15in}
\noindent $^f$ $F_0=10^{-11}$ erg cm$^{-2}$ s$^{-1} $
%\vspace{-0.15in}
%\noindent
%\hspace{0.6in} 
%\noindent

These are only qualitative criteria but for the IC model one can give some
{\it quantitative estimates}.  At a redshift $z\equiv Z-1$ the CMB energy
$u\propto Z^4$ and the critical frequency $\nu_{c,{\rm IC}}\propto Z$.
As a result the ratio of IC to synchrotron fluxes
\be\label{ratio}
{\cal R}=F_{\rm HXR}/F_{\rm radio}\propto Z^{2+\Gamma }/B_\perp^{\Gamma},
\ee
so that for an observed radio flux one
would like to choose clusters with the lowest magnetic field $B$
and the highest redshifts.
Unfortunately the $B$ field strengths in
most of the diffuse radio emitting clusters are not known (Coma is an
exception) so that one
must rely on some theoretical arguments to estimate the value of $B$.
If  we assume some proportional relation (\eg {\it equipartition}) between
the energies of the magnetic field and non-thermal electrons

\be\label{equip}
{\cal E}_e=N_\tot{p-1 \over p-2}\g_\min mc^2=\zeta {B^2\over 8\pi}{4\pi R^3\over 3},
\ee
where $R=\t d_A/2$ is the radius of the (assumed) spherical
cluster with the measured angular diameter $\t$ and
angular diameter distance $d_A(Z)$, and equipartition with electrons is equivalent to
$\zeta = 1$.

From the three equations  (\ref{nufnu}), (\ref{ratio}) and (\ref{equip})
we can determine the three unknowns $B, {\cal E}_e$ (or $N_\tot$) and $F_{\rm HXR}$
purely in terms of $\zeta, \g_\min$, and the observed quantities
(given in Table 1) $z, \t$ and the radio flux

\be\label{flux}
\nu F_{\rm radio}(\nu) = cr_0^2N_\tot A_\sy(p)(p-1)\g_\min^{p-1} (B^2/8\pi)(Z\nu/\nu_{\crit,sy})^{2-\Gamma}/(4\pi d^2_L(Z))
\ee
where $d_L=d_AZ^2=Zr(Z)$ is the luminosity distance to a source with 
the co-moving coordinate $r(Z)$. The result is%
\footnote{Here we have set the Hubble constant $H_0=70$ km Mpc$^{-1}$ s$^{-1}$,
the CMB temperature $T_0=2.8$ K, and the radio frequency $\nu = 1.4$ GHz. 
In general $B^{2+\Gamma} \propto H_0\nu^{\Gamma-1}, N_\tot\propto H_0^{-3}$
and $F_{\rm HXR}\propto H_0^2T_0^{2+\Gamma}$. We have also assumed an 
isotropic distribution of the electron pitch angles and set  $B=B_\perp(4/\pi)$.}

\be\label{B}
(B/\mug)^{\Gamma+2}=0.20 \, \zeta^{-1} \left({F(1.4 \, {\rm GHz}) \over 
{\rm Jy}}\right)\left({5' \over \t}\right)^3\left({10^4 \over \g_\min}\right)^{2\Gamma-3}
{Z^{3-\Gamma}\over r(Z)},
\ee
\be\label{Ntot}
N_\tot = 2.3\times 10^{65} \, {\Gamma-3/2\over \Gamma-1}\zeta\left({10^3 \over \g_\min}\right)\left({\t \over 5'}\right)^3
\left({B \over \mug}\right)^2\left({r(Z) \over Z}\right)^3
\ee
and

\be\label{HXRflux}
\epsilon F_{\rm HXR}(\epsilon)=0.034\times F_0\left({N_\tot \over 10^{65}}\right)
\left({10^4 \over \g_\min}\right)^{2\Gamma-4}\left(
{\epsilon \over 5.9 \, {\rm keV}}\right)^{2-\Gamma}\left({Z \over r(Z)}\right)^2, 
\ee
where we have defined $F_0\equiv 10^{-11}$ erg cm$^{-2}$ s$^{-1}$.
Note also that in all these expressions one may use the the substitution 
$R=3.39 \, {\rm Mpc} \, (\t/5') \, r(Z)/Z$.  

\begin{figure}[htbp]
\leavevmode\centering
\plotone{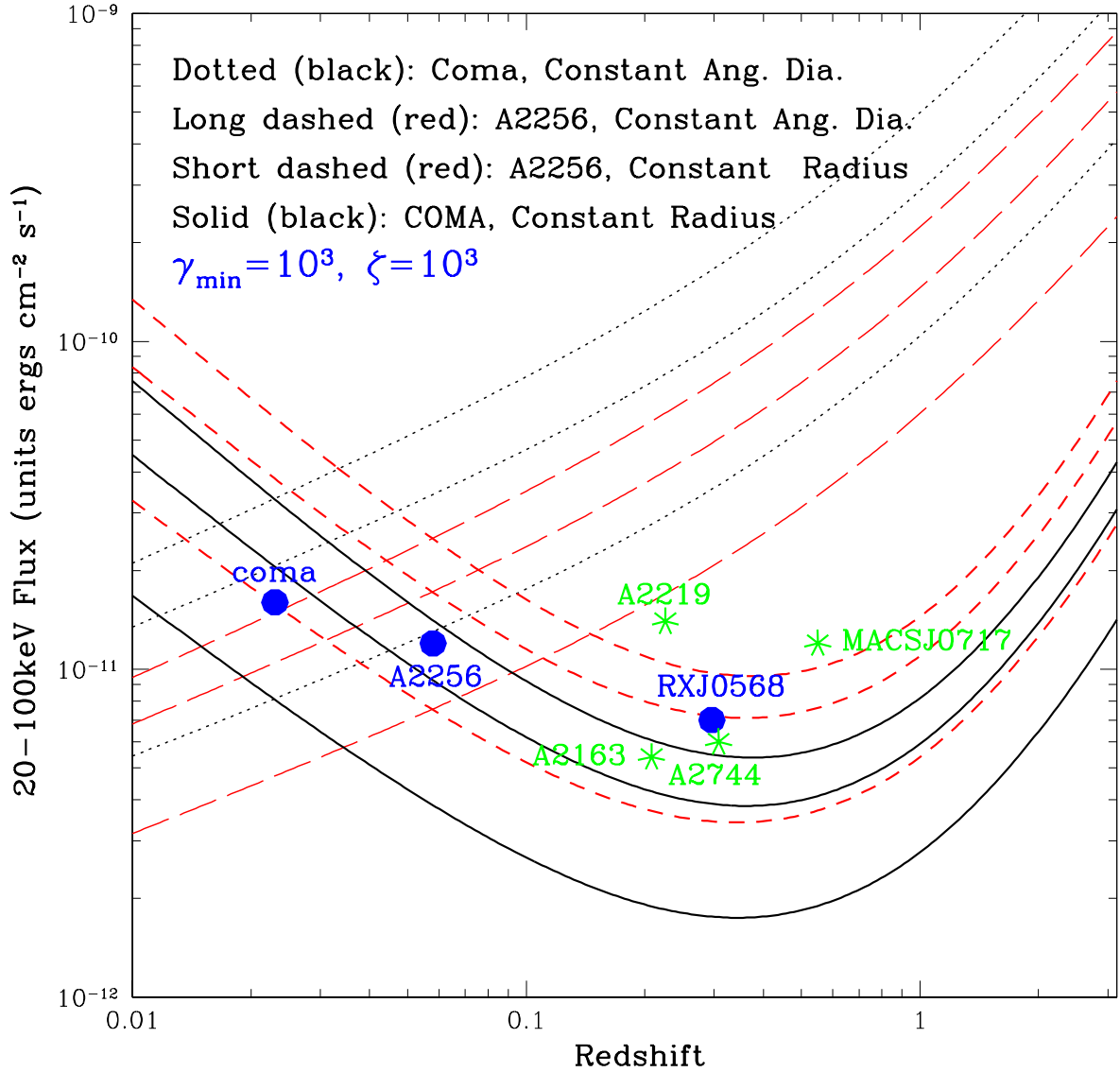}
\caption{\footnotesize Predicted variations of the HXR flux
with redshift assuming a constant metric
(solid and dashed) or   angular (dotted and long-dashed) diameters, 
using COMA (black and) and A2256 (red) parameters 
assuming $\zeta=\g_\min=10^3$.
In each group the IC photon spectral
index  $\Gamma =  2.25, 2.0, 1.75$, from top to bottom. Filled circles based 
on observations and stars based on the estimates given in Table 1.
}
\label{Fluxes}
\end{figure}

To obtain numerical estimates of the above quantities in addition to the 
observables $F_{\rm radio}, \t$ and redshift $z$ we need
the values of $\zeta$ and $\g_\min$. Very little is known about  
these  two parameters and how they may vary from cluster to cluster.
From the radio observations at the lowest frequency we can set an upper 
limit on $\g_\min$; for Coma \eg
$\g_\min<4\times 10^3 (\mug/B)$. We also know that the cut off 
energy cannot be too low because that will require excessive amount of 
energy (for $p>3$) which will go into heating of the ICM gas via Coulomb
collisions. A conservative estimate will be $\g_\min\sim 10^3$.
Even less is known about $\zeta$.
The estimated values of the magnetic fields $B$ for the simple case of 
$\Gamma=2$, equipartition (\ie $\zeta=1$) and low energy cut off 
$\g_\min=10^3$ are given in the 7th column of Table 1. As expected
these are of the order of a few $\mug$;  for significantly stronger field, 
the predicted HXR fluxes will be 
below what is detected (or even potentially detectable). 
For $\Gamma=2$ the magnetic field $B\propto (\zeta\g_\min)^{-1/4}$ and  
$F_{\rm HXR}\propto (\zeta\g_\min)^{1/2}$ so that for sub-$\mug$ fields and 
$F_{\rm HXR}\sim F_0$ we need $\zeta\g_\min\sim 10^6$. Assuming $\g_\min= 10^3$
and $\zeta=10^3$ we have calculated the expected fluxes integrated in the 
range of $20-100$ keV (which for $\Gamma=2$ is equal to 
$1.62 \times (20 \, {\rm keV} \, F_{\rm HXR}(20 \, {\rm keV})$)
shown on the last column of Table 1.
The variation of this flux  
with redshift based on the observed parameters, 
$\t$ and $F_{\rm radio}(\nu=1.4 \, {\rm GHz})$
of Coma and A2256 are plotted in Figure \ref{Fluxes} 
for three values of $\Gamma=1.75, 2.0$ and 2.25 ($p=2.5, 3$ and 3.5) 
and assuming a constant metric radius $R$ which a reasonable assumption.
We also plot the same assuming a constant angular diameter.
This could be the case due to observational selection bias 
if most diffuse emission is near the resolution of the telescopes.  
These are clearly uncertain procedures and can give only semi-quantitative
measures.  However, the fact that the Coma and A2256 have the highest
fluxes, and that our observations of 1E0657-56 described below, yield a flux
close to our predicted value is encouraging. Clearly, all the other clusters 
in Table 1 are similarly promising candidates for future HXR observations.

\section{OBSERVATIONS AND ANALYSIS}
\label{obs}

\subsection{Description of Observations}

Cluster 1E0657-56 was observed with \textit{RXTE }during 95 separate
pointings from 2002 August 19 to 2003 April 3, collecting approximately
400,000 seconds of data. After filtering the PCA data by following the
standard selection criteria, we obtained a total of 309,630 seconds of good
PCA data. During the campaign, only PCU 0 and PCU 2 were on. Since PCU 0 had
lost its propane guard layer, in order to ensure that we have the best
possible calibration, we used only the PCU 2 detector. The selection
criteria involved excluding data obtained when the detector pointing was
more than $0.02%
%TCIMACRO{\U{b0}}%
%BeginExpansion
{{}^\circ}%
%EndExpansion
$ off the source, excluding data obtained when the Earth elevation angle was
less than $10%
%TCIMACRO{\U{b0}}%
%BeginExpansion
{{}^\circ}%
%EndExpansion
$, and excluding times near passage through the South Atlantic Anomaly which
caused large variations in the count rate.

The amount of good time for each of the two HEXTE detector clusters is about
one-third that for the PCA data: 105,450 seconds for Cluster A and 105,510
seconds for Cluster B. The ratio of HEXTE net time to PCA net time is
reasonable, since each HEXTE cluster spent half of the time for background
measurement and a fraction of the time was lost due to electronic dead time
caused by cosmic rays.

The PCA background was estimated using the 
{\tt ftool} {\tt pcabackest} (v 3.0) with the
``L7/240'' model provided by the \textit{RXTE }GOF (Guest Observer
Facility).  The estimator program used the model known as 
${\tt pca\_bkgdcmfaint17\_eMv20051128.mdl}$ as the 
background model, which in turn 
took into consideration the instantaneous particle-induced intrinsic 
instrument background and activation - mainly due the SAA (South
Atlantic Anomaly) passages - as well as the the cosmic X-ray background.  
We note here that the above background model, available from HEASARC at 
NASA's GSFC (Dr. Craig Markwardt, priv. comm.) 
allows robust estimate of background 
up to $\sim 30$ keV even for faint sources such as 1ES0657-56.   

We used the top layer of PCU 2 in the $ 3 - 30$ keV range.  
The mean background subtracted PCU2 count rate
was $1.418\pm 0.007$ count s$^{-1}$ in the top layer over 
this band.  For the response matrix, we used the 
standard PCA response matrix, generated via the 
${\tt pcarsp}$ software tool (ver. 10.1), 
as appropriate to the beginning of the observations (2002 August);  
while the \textit{RXTE } PCA response matrix changes with time, the change is
gradual, and the values of spectral fits obtained using the 
matrix appropriate for the end of the observations 
(2003 April) showed no discernible difference.  For HEXTE, the
background subtracted count rates over the 20 $\sim $ 70 keV energy band
were $0.0103\pm 0.0227$ count s$^{-1}$ and $0.0700\pm 0.0192$ count s$^{-1}$ 
for HEXTE detector clusters A and B, respectively;  above 70 keV, the
source was clearly not detected.  We used the standard redistribution files, 
{\tt xh97mar20c\_pwa\_64b.rmf} and {\tt xh97mar20c\_pwb013\_64b.rmf}, 
with the standard HEXTE effective area files available through HEASARC.  

While the \textit{RXTE } data clearly detected the cluster, 
an inclusion of higher 
spectral resolution data allows a better constraint on the temperature 
of the thermal component and thus better constraints on the nature of 
the non-thermal emission.  To this end, we considered the published 
\textit{ASCA } data, but because the best 
statistical accuracy was provided 
in the \textit{XMM-Newton } data, we extracted and analyzed 
archival spectra collected during the \textit{XMM-Newton } pointing 
in 2000 October 20-21.  
We reduced the data using a procedure described 
in Andersson \& Madejski (2004), but here, we  used the 
${\tt XMMSAS\_20050815}$ release of the \textit{XMM-Newton } 
data analysis software.  
We screened the data against any obvious 
flares;  this meant any segments of data with the total count rate greater 
than 12 counts s$^{-1}$ for {\tt pn}, and 4.5 
and counts s$^{-1}$ for either 
MOS.  After cleaning, the effective exposure times are 24,916 s 
for the MOS1 data, 24813 s for the MOS2 data, and 21,704 s for the 
{\tt pn} data.  
We note that the detailed analysis of the \textit{XMM-Newton } data for 
this cluster - and in particular, the analysis of the spatial structure - 
will be presented in Andersson et al (in preparation).  
For all three \textit{XMM-Newton } detectors, we extracted 
source counts from a region of $4'$ radius around the centroid of the count 
distribution.  For background, we used an annulus with inner and outer radii 
of $8'$ and $12'$, with all obvious point sources removed.  
We also removed the data corresponding to the Cu K line in the {\tt pn} 
camera, 
corresponding to 7.9 to 8.2 keV spectral range.  The data were 
grouped to include at least 40 counts per bin for the {\tt pn} data files, 
and 25 counts per bin for the MOS data files.  
We used the standard instrument resolution 
matrices and effective area files as released 
in the SAS version as above.  In the subsequent 
analysis, we restricted the two MOS and {\tt pn} 
data to the 1 - 10 keV range, to 
avoid any residual cross-calibration problems between the three instruments.  

\subsection{Analysis and Spectral Fitting}
\label{spec}

\bigskip

\subsubsection{\textit{RXTE }Data}
\label{rxte}
We first present an analysis of the \textit{RXTE } observations and then
a joint analysis of the \textit{RXTE} and
\textit{XMM-Newton} data.  In all subsequent 
fits we fix the redshift at 0.296.  The models 
consist of a {\tt mekal} thermal
emission component, where we also allow a second component, either thermal, 
or a power law.  We first use a 
single thermal {\tt mekal} 
model to fit the \textit{RXTE} data. The model includes
a cold absorption due to neutral gas in the line of sight 
in our Galaxy. Since the \textit{RXTE} data alone do not constraint the
Galactic column density, we fix this parameter at $4.6\times 10^{20}$ 
cm$^{-2}$, the value from radio measurements, but also consistent with the 
best fit to \textit{ASCA} data (Liang et al. 2000).  We note that all
errors quoted correspond to 90\% confidence limits.  

The best fit temperature under an assumption of a simple, 
one-temperature {\tt mekal} model 
is $12.1\pm 0.4$ keV.  This is in moderate agreement with the value 
$14.5_{-1.7}^{+2.0}$ keV determined from a joint fit to the \textit{ROSAT }
and \textit{ASCA } data (Liang et al. 2000), and the value $14.8_{-1.2}^{+1.7}
$ keV from \textit{Chandra }(Markevitch et al. 2002). The metal abundance is
$A=0.16 \pm 0.04$ Solar, which is moderately lower than the value 
$A=0.33 \pm 0.16$ from the joint \textit{ROSAT } and \textit{ASCA } data but is
in good agreement with the \textit{Chandra } value $0.11\pm 0.11.$  The
best-fit model yields the $\chi ^{2}$ of $114$ for $98$ degrees of
freedom. The ratio of the \textit{RXTE } data to the best-fit {\tt mekal} 
model indicates systematic upward rise above 10 keV, 
suggesting that an extra component may be required.  If we adopt this 
secondary component to be a power-law, then $\chi ^{2}$ is $102$ (96 degrees
of freedom), with poorly determined index ($\Gamma > 2$). 
The 90\% confidence interval of the power-law flux in the 20--100 keV 
range 
is $\left(0.3\pm 0.2\right)F_0$.  
We thus conclude that even from the \textit{RXTE } 
data alone, we have a marginal 
but suggestive detection of hard X-ray flux in 1E06567-56.  

\subsubsection{\textit{XMM-Newton } Data}

The \textit{XMM-Newton } data extracted as above were first fitted to a model 
including absorption due to neutral gas with Solar abundances 
and a thermal ({\tt mekal}) plasma.  Since
the absolute calibration of the three \textit{XMM-Newton } detectors 
might vary, we allowed the respective normalizations not to be tied 
to each other.  The best fit 
absorption was close to the value inferred from the radio data 
of Liang et al. (2000), with the column $2.8 \pm  1.0 \times 10^{20}$ 
cm$^{-2}$, but here, inclusion of the MOS data below 1 keV altered the 
resulting best fit column to $5 \pm  0.5 \times 10^{20}$ cm$^{-2}$,
We concluded that the column derived from the \textit{XMM-Newton } data 
is consistent with  $4.6 \times 10^{20}$ 
cm$^{-2}$ cited by Liang et al. (2000), and used that value 
in all subsequent fits.  
The temperature of the {\tt mekal} component was $12.0 \pm 0.5$ keV, 
and elemental abundances $0.24 \pm 0.04$ Solar, consistent with all 
data sets above.  The fit is acceptable, at $\chi^2 = 1367$ for 1408 d.o.f.  

\begin{center}
\center{{\large {\bf Table 2}}}
\vspace{-0.16in} \center{\bf PARAMETERS FROM SPECTRAL FITTINGS}

{\small {%
\begin{tabular}{|l|c|c|c|c|}
\hline\hline
Data Set & Parameter & Single Thermal & Double-Thermal & Thermal+Power Law
\\ \hline
& $\left( n_{H}/10^{20}{\Large \ {\rm cm}}^{-2}\right) $ & $4.6^f$ & -- & $4.6^f$ \\
& $kT_{1}$\ $\left( {\rm keV}\right) $ & $12.1\pm 0.4$ & -- & $11.7\pm 0.5 $ \\
& $kT_{2}$\ $\left( {\rm keV}\right) $ & -- & -- & -- \\
\textbf{RXTE} & Abundance (Solar) & $0.16\pm $\ $0.04$ & -- & $0.25\pm 0.08$ \\
& Photon Index & -- & -- & $> 2$ \\
& ${F_{20 keV}^{100 keV}}/F_{0}$ & -- & -- & $0.3\pm 0.2$\\
& $\chi ^{2}/dof$ & $114/98$ & -- & $102/96$ \\ 
\hline
& $\left( n_{H}/10^{20}{\Large \ cm}^{-2}\right) $ & $2.8\pm 1.0$ & $4.6^f$ & $4.6^f$ \\
& $kT_{1}$\ $\left( {\rm keV}\right) $ & $12.1\pm 0.2$ & $10.1 \pm 0.9$ & $ 11.2 \pm 0.8 $ \\
\textbf{RXTE and} & $kT_{2}$\ $\left( {\rm keV} \right) $ & -- & $50 \, (> 30)$ & -- \\
\textbf{XMM} & Abundance (Solar) & $0.19\pm 0.03$ & $0.19\pm 0.03$ & $0.22\pm 0.04$ \\
& Photon Index & -- & -- & $1.6\pm 0.2$ \\
& $F_{20 keV}^{100 keV}/F_{0}$ & -- & $0.5\pm 0.3$ & $0.5\pm 0.3$
\\
& $\chi ^{2}/dof$ & $1483/1508$ & $1471/1506$ & $1464/1506$ \\ \hline\hline
\end{tabular}%
}}

\end{center}

\noindent $F_{0}=10^{-11} \, {\rm erg\, cm}^{-2}\,{\rm s}^{-1}$\\
$^f$ denotes parameter fixed at the given value\\
%\end{center}

\subsubsection{Joint \textit{RXTE } and \textit{XMM-Newton }Data}
\label{combined}

In the analysis below, we include both the \textit{RXTE } and 
\textit{XMM-Newton } data, 
since the very good effective area of the instrument 
coupled with good spectral 
resolution of its detectors 
allows tighter constraints on spectral parameters for 
the emission detected below 
$\sim 10$ keV, and thus mainly on the thermal component.  However, the only
reliable approach here is to perform the spectral fitting simultaneously.  

The joint analysis of the \textit{RXTE } and \textit{XMM-Newton } 
data provide more evidence of the need for a secondary component. 
We use the \textit{XMM-Newton } and \textit{RXTE } data over the bandpasses 
as above;  we allow the normalization of the \textit{RXTE } 
instruments to be different 
than that for the \textit{XMM-Newton } instruments, which in turn are allowed 
to vary among themselves.  A single isothermal ({\tt mekal}) fit yields an 
adequate fit, with $\chi^2$ of 1483 for 1508 degrees of freedom. 
The hydrogen column density $n_{H}$ is $\left( 2.8\pm
1.0\right) \times 10^{20}$ cm$^{-2}$, which is marginally lower than the
Galactic value of Liang et al. (2000) (see the discussion above);  in the 
subsequent fits we adopt the value of Liang et al. (2000) of 
$4.6 \times 10^{20}$ cm$^{-2}$.  Regardless of the exact 
value of absorption, the plasma temperature is $12.1\pm 0.2$
keV, and elemental abundances are $0.19\pm 0.03$ Solar.  
We note that the true errors 
on those quantities are only approximate, since as pointed out by Markevitch 
et al. (2002;  2004), this cluster 
shows some temperature structure, while we use 
an average temperature.  

Adding a power-law component to the
isothermal model improves the $\chi ^{2}$ value, which is now 
1464/1506 d.o.f.   Such change in $\chi ^{2}$ - with addition 
of two parameters - is very significant (at more than 99.9\%).  
The temperature in this case decreases 
to $11.2\pm 0.8$ keV.  The photon index is $1.6\pm
0.3$ and the non-thermal flux in the 20--100 keV energy band is 
$\left(0.5\pm 0.2\right)F_0.$  
In the context of statistical significance of raw counting rates, 
at energies above 20 keV - where the power law model dominates - 
the excess (over the value predicted by the thermal model) 
corresponds to $\sim 0.8 \sigma$ and $2.8 \sigma$ 
for HEXTE Cluster A and B, respectively.  
For the purpose of illustration, we  
show the unfolded spectrum of the data fitted to a single-temperature thermal 
model in Figure 2, while Figure 3 shows the 
confidence levels of the two fit parameters, photon index vs. 
the flux of the non-thermal component, for our preferred two-component model. 

%\protect\bigskip

\begin{figure}[htbp]
\begin{center}
\includegraphics[angle=270, width=130mm]{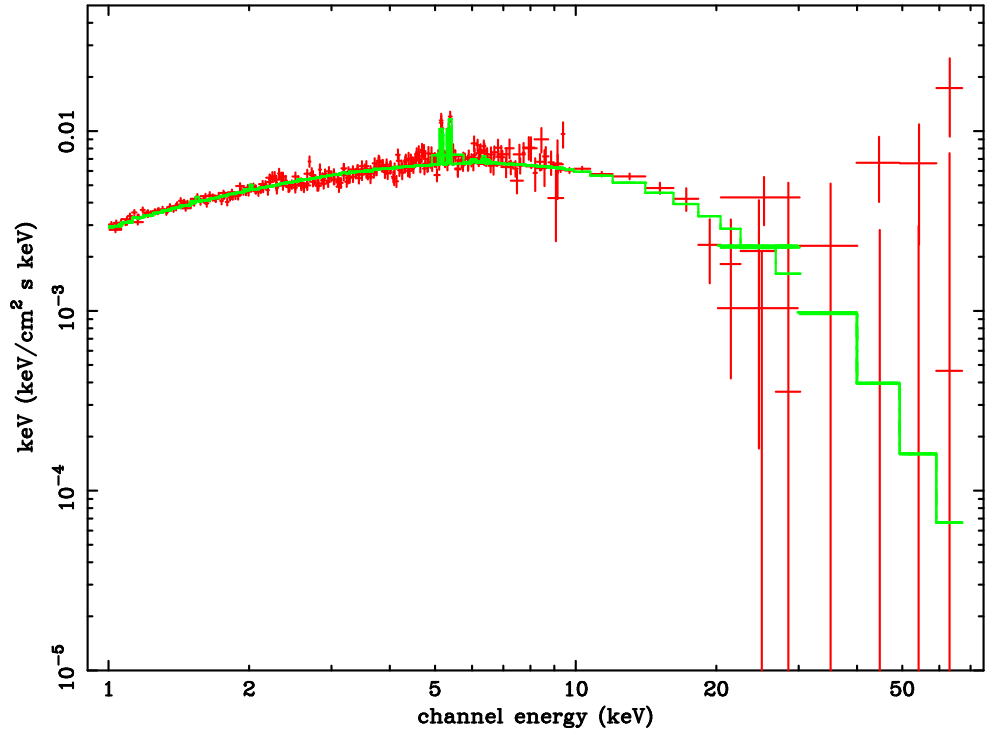}
\end{center}
\caption{Unfolded spectrum 
of the cluster 1E0657-56, where the \textit{XMM-Newton } 
and \textit{RXTE } data were fitted with a single-temperature 
thermal ({\tt mekal}) model, plotted here as the solid-line histogram.  
For the purpose of clarity of the plot, prior to plotting, 
the \textit{XMM-Newton } data were scaled by a cross-normalization 
factor, close to unity, to agree with the \textit{RXTE } data.  }
\end{figure}

Alternatively, this secondary, hot component can be modeled via another 
thermal plasma spectral component.  Using this parametrization (and assuming 
that only the temperature and normalization differ from the lower $T$ 
component), we 
inferred its temperature to be $50 \, (> 30) $ keV.  The $20 - 100$ keV flux
of this component is also 
$0.5 \pm 0.2 \times F_0$.  Now the lower $T$ component 
has a temperature of $10.1 \pm 0.9$ keV.  While we cannot clearly exclude this 
interpretation on the quality-of-fit grounds ($\chi^2$ is 1471 
for 1506 d.o.f.), we argue in Sec. 4 that the power law spectral shape 
provides for a more viable interpretation of the secondary component.  

\begin{figure}[htbp]
\begin{center}
\includegraphics[angle=270, width=130mm]{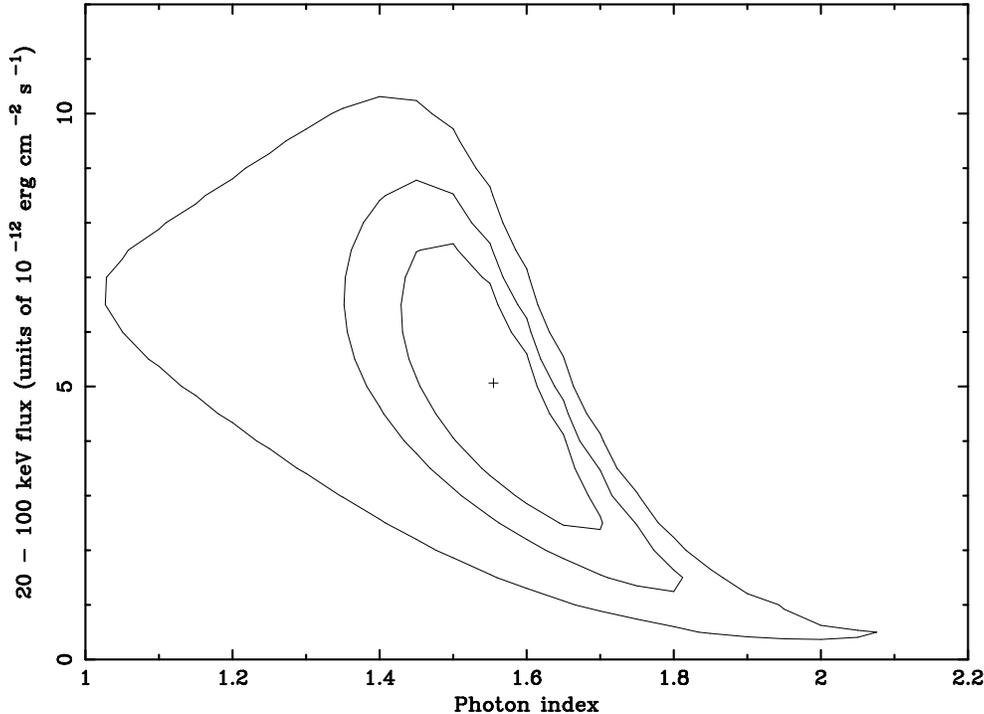}
\end{center}
\caption{Confidence regions (corresponding to $\chi^2_{min} + 2.3, 4.6,$ and 
9.2) on the best fit parameters for the spectral fits to the 
joint \textit{XMM-Newton }, \textit{RXTE } PCA, and HEXTE 
data for cluster 1E0657-56 for the photon index of 
the non-thermal X-ray component and its 20 - 100 
keV flux (in units of $10^{-12}$ erg cm$^{-2}$ s$^{-1}$).  
See the text for details.  }
\end{figure}

\section{SUMMARY AND DISCUSSION}
\label{disc}

Radio observations of diffuse emission from ICM show presence of 
non-thermal activity in many clusters, especially in those with high 
SXR emitting gas temperatures (and luminosity) and showing recent merger 
activity.  A list of such clusters is shown in Table 1. Detections of
HXR emission exceeding the levels expected from the thermal gas have 
been reported by two groups using different instruments. This enables a  
quantitative investigation of the
nature and origin of the non-thermal activity. The radio observations alone
indicate presence of extreme relativistic electrons ($\g\sim 10^4$). 
Assuming equipartition between the relativistic electrons and the 
magnetic field we estimate a volume-averaged magnetic field value 
in the range of 0.5 to 2 $\mug$ (see Table 1), 
in rough agreement with (line-of-sight averaged)
field strengths deduced from Faraday rotation measurements.
In the case of Coma the magnetic field deduced from Faraday rotation 
of $2-3 \, \mug$, implies conditions far from equipartition if we assume 
a homogeneous source;  the field energy density 
is about 25 times larger than that of the electrons.  This 
discrepancy can be resolved by an inhomogeneous model; \eg the two phase 
model proposed by Brunetti \ea (2001) that implies a magnetic field profile, 
or if  there was 25 times more 
energy in nonthermal protons than electrons.

If the HXR radiation is due to IC scattering of CMB photons by the the same 
electrons that produce the synchrotron radio emission,
which we have argued to be the simplest scenario, such high magnetic files
would imply IC fluxes in the $20-100$ keV range of about $10^{-2}F_0$ while
both Coma and A2256 show fluxes higher than $F_0$. Production of HXRs at 
this level require magnetic fields of about $0.1 \, \mug$ (or partition 
parameter $\zeta\sim 10^3$) which raises the required 
energy of the non-thermal electrons (${\cal E}_e\propto \zeta^{1/2}$). 
For example for Coma and A2256 the observed HXR fluxes
require ${\cal E}_e= 2\times 10^{-13}$ and 
$1.5\times 10^{-12}$ erg $\cc$, respectively, which are
comparable to the energy density of the CMB ($4.6\times 10^{-13}$ erg $\cc$). 
Assuming similar conditions
for other clusters, several years ago we identified 1E0657-56 and A2219 as 
possible candidate for having detectable HXR fluxes. 
% We proposed \textit{RXTE } observations of both clusters (the second  with 
% Y. Rephaeli as the CoI) and were granted observing time for the former
% which has much higher redshift than those detected previously.
Here, we have described these observations of 1E0657-56 and the results of 
our spectral fitting to these data alone and to the \textit{RXTE } 
data combined with the SXR data from \textit{XMM-Newton}. 
In the future, attempts should be made to observe 
other clusters in Table 1, and any newly discovered cluster with a 
relatively high temperature and redshift, an ample sign of substructure 
and/or mergers, and, of course, with a strong diffuse
radio emission. A low value of magnetic field indicated by Faraday rotation 
measurements or estimated via equipartition argument will also be good 
justification for HXR observations.  

We have shown that a thermal component plus a non-thermal tail 
described as a power law with spectral index 
$\Gamma=1.6$ containing $\sim 0.5 F_0$ of the $20-100$ keV flux provides an 
acceptable fit. This flux is fortuitously close to the value of $0.52 \, F_0$
we estimate by scaling the Coma and A2256 observations. (Note that the 
equipartition (\ie $\zeta=1$) magnetic field and flux for this cluster 
would be $\sim 1 \, \mug$ and $0.02 \, F_0$.) 
This agreement is somewhat encouraging so that we believe that 
the chances  are even better for  detection of HXR emission from Abell 2219, 
a good candidate  at a moderate redshift,  and a 
recently discovered cluster, MACJ7017, with even higher $z$ than 1E0657-56. 
As is the case for Coma and Abell 2256, these clusters also require high 
nonthermal electron energy densities; 
${\cal E}_e= [1.0, 1.8, 8.5]\times 10^{-13}$
erg $\cc$
for Abell 2219, 1E0657-56 and MACSJ0717, respectively. 
These are even closer to the CMB
densities at their respectively higher redshifts. 
In general the required electron energies are 
about 1 to 10 \%  of the energy density
of the thermal gas of few times $10^{-11}$ erg $\cc$ 
which is also comparable to 
the gravitational potential energy density of the cluster. 
This means a very efficient
acceleration process.

Regarding the spectral fits to 1E0657-56, as shown in Table 2, 
we also obtain an acceptable fit with a double temperature
thermal model with a second component with $kT=50$ keV and a normalization
implying a volume emission measure ($EM$) for this component 
which is a significant 
fraction ($\sim 10\%$) of that of the lower temperature component.
Indeed Markevitch (2005) finds evidence for such a high $T$ component 
behind the shock near the bullet based on the recent Chandra 
Observatory data.  However, this has a much smaller
$EM$ than is required for fitting the data with a two temperature 
model. In fact a high $T$ component with such a large $EM$ would have 
been easily detected by Markevitch (2005). 
Clearly more and deeper observations of the 
clusters in Table 1 can clarify this situation considerably. 
In particular, 
deep observations with imaging HXR instruments -- such
as those with the proposed {\sl NuSTAR} satellite -- which can  provide 
information about the spatial distribution of the HXRs  would be very valuable.

In spite of the uncertainties about the exact character  of the 
observed radiation spectra and the emission mechanism, and in spite of the
meagerness of the
data, the acceleration mechanism of electrons can  be constrained
significantly.  The lifetimes of electrons with energies in the range
$200 \, {\rm keV} \leq E \leq 200 \, {\rm GeV}$ are longer 
than the crossing time, $T_\tr\sim
3\times 10^6$ yr.  Therefore, these electrons will escape the cluster and
radiate most of their energy outside of the cluster unless 
there exists some scattering
agent with a mean free path $\lambda_\scat\sim 1$ kpc to trap the electrons
in the ICM for at least a timescale of $T_\esc=(R/\lambda_\scat)T_\tr\sim
3\times 10^9$ yr, for cluster size $R\sim 1$Mpc.  
Turbulence can be this agent and
can play a role in stochastic acceleration directly, or indirectly in
acceleration by shocks, presumably arising from merger events.  
Several
lines of argument point to an ICM which is highly turbulent, and there has
been considerable discussion of these aspects of the problem in the recent
literature (see, e.g., Cassano \& Brunetti 2005).  
These aspect are explored in P01, but the upshot of which is
that {\it we require injection of high energy electrons, 
presumably from galaxies
and AGNs, and that the acceleration process is episodic on time scale of $\sim
10^8$ to $10^9$ yr.} Whether the electrons are injected 
directly by past and current AGN 
activity, or are produced by the interaction with the 
thermal gas of cosmic ray 
protons escaping the galaxies would be difficult to determine now. There are 
however constrains on both these scenarios (see \eg Blasi 2003).  

\acknowledgments

The work was supported by NASA's ADP grant NNG04GA66G-NCX,
by Chandra
grants GO1-2113X and GO4-5125X from NASA via Smithsonian
Astrophysical Observatory, and by the Department of Energy
contract to SLAC no. DE-AC3-76SF00515. 
We acknowledge the help in reducing the \textit{XMM-Newton} 
data by Mr. Karl Andersson, and Dr. Craig Markwardt for providing us 
with the updated \textit{RXTE } PCA background information.

{}

\end{document}